\documentclass[12pt]{article} 
\usepackage{amssymb}
\usepackage{graphicx}
\begin{document} 
\begin{center}
{\large \bf The multilayer structure of protons } 

\vspace{0.5cm}                   
{\bf I.M. Dremin}

\vspace{0.5cm}                       
         Lebedev Physical Institute, Moscow 119991, Russia\\

\end{center}

\begin{abstract}
It is argued that the dynamics of the elastic scattering of high-energy
protons at intermediate transferred momenta changes with the energy increase.
It evolves from the multiple scattering at the external layer for energies about
10 GeV to the double scattering at the two subsequent layers within the
colliding protons for energies about 10 TeV. The problem of the unitarity is
considered in this context.
\end{abstract}


\vspace{0.5cm}                       


\vspace{0.5cm}                       

The collaboration TOTEM published the data on the differential cross section
of elastic scattering of protons at the energy 13 TeV \cite{tot, cso}. 
It decreases approximately exponentially (with slight oscillations as seen
in the insert in Fig. 1)
\begin{equation}
d\sigma /dt \propto \exp [-B\vert t \vert]; \;\; B \approx 20.4 \; GeV^{-2}
\label{diff}
\end{equation}
at small transferred momenta 
$0.04<\vert t \vert =2p^2(1-\cos \theta)<0.2\;$ GeV$^2$ (where
$p$ is the momentum of colliding protons and $\theta $ is their scattering 
angle). The exponent $B$ shows the size $R$ of the scattered protons
$(B\approx R^2).$ It increases logarithmically with the energy increase in
accordance with many theoretical models. The dip at $\vert t\vert =0.47\;$GeV$^2$
is usually interpreted as a consequence of zero value of the imaginary part of
the amplitude at that point. 

More surprising is the behavior of the cross section at somewhat larger
transferred momenta $0.7<\vert t\vert <3.83$ GeV$^2$ (see Fig. 8 and Tables 9 
and 10 in \cite{tot}). It shows also the
exponential decrease albeit with the much smaller exponent
\begin{equation}
d\sigma /dt \propto \exp [-C\vert t \vert]; \;\; C \approx 4.3 \; GeV^{-2},
\label{inter}
\end{equation}
In analogy with the spatial interpretation of the exponent $B$, one is tempted 
to assume that this exponent gives a hint at a new deeply
positioned layer inside the protons \cite{dr18}. In this regard, it reminds
of the Rutherford discovery of the nuclei inside atoms.

\begin{figure}
\centering
\includegraphics[width=12cm, height=8cm]{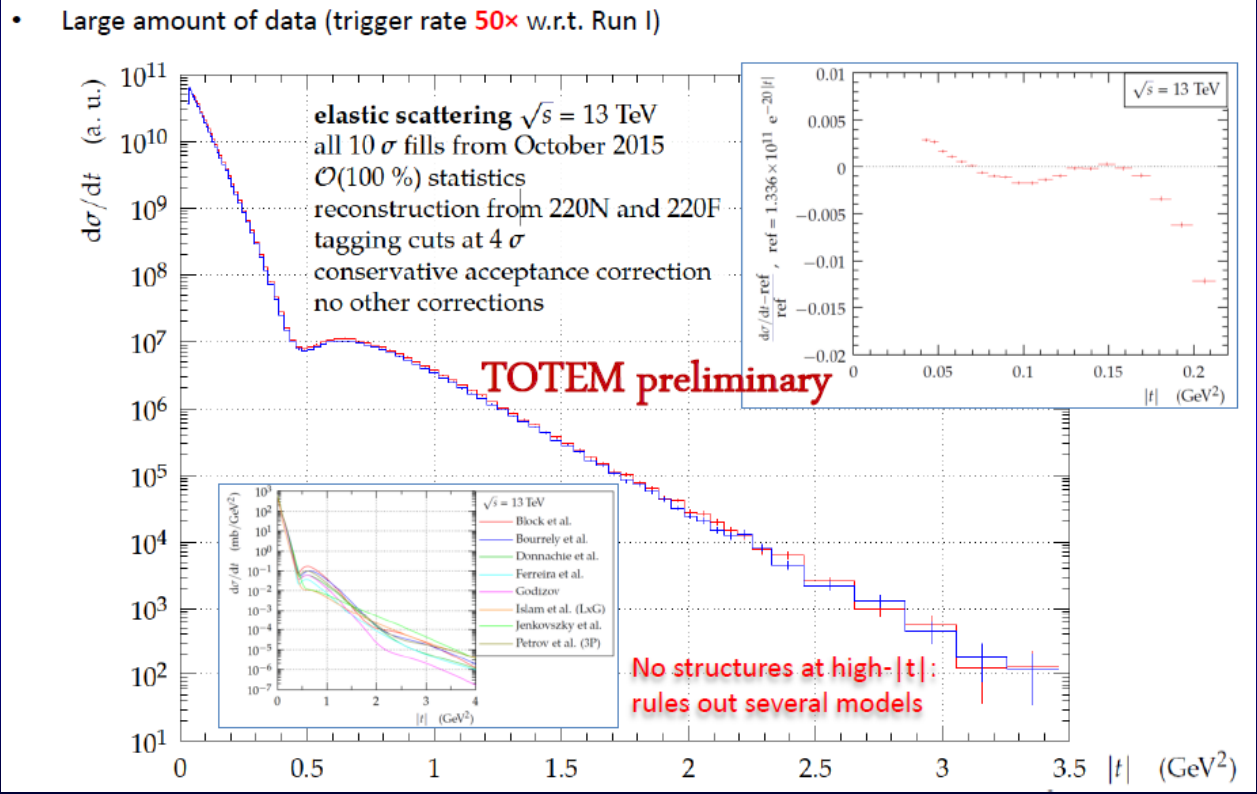}

\caption{The differential cross section of elastic scattering of protons at 
13 TeV shown in \cite{cso} and discussed in \cite{dr18, dr19}. The upright insert
demonstrates slight oscillations within the diffraction cone. Another insert
contains some theoretical predictions outside the diffraction cone.}
\label{fig1}
\end{figure}

From the early days of Yukawa's prediction of pions, the spatial size of
protons was ascribed to the pionic cloud surrounding their centers. The pion
mass sets the scale of the size in the order of 1 fm=10$^{-13}$ cm. Numerous
low-energy experiments using different methods confirmed this estimate with
values of the proton radius ranging from 0.84 fm to 0.88 fm. This 5$\%$
difference has been named the "proton radius puzzle". The very external shell
of a proton is usually described as formed by single virtual pions as the
easiest color-neutral particle constituents. That is why the one pion exchange 
model was first proposed \cite{drch} for the description of the peripheral
interactions of hadrons. It initiated numerous multiperipheral (multireggeon)
models. 

The internal layers of protons must contain more massive constituents and be
responsible for scattering at larger angles. Their study at larger transferred 
momenta asks for heavier exchanged objects and higher energies. As a particular
example, let us mention the model of three-layered protons proposed in Refs
\cite{isl, isku}. The internal layer is ascribed to $\omega $-exchanges, and
the central layer consists of three quarks with a junction. The special 
formfactors are used for these layers with 17 adjustable parameters. However,
the fits of experimental data obtained in the model are still not very 
successful so that it can be considered just as one of theoretical attempts.
The change of the pressure inside the proton layers from the attractive one
at the periphery (strongest at about 0.9 fm) to the repulsive one at the central 
regions (less than 0.6 fm) was found in Ref. \cite{123}. Some intriguing 
features of the developing hollow at small impact parameters were noticed 
recently
\cite{dr17, cps} in the spatial image of proton interactions profile at 13 TeV.
That could also be a signature of some new structures revealed at high energies.

The layer structure of protons can be at the origin of different behavior of
their scattering outside the diffraction cone with the rise in collision 
energies. At the energies about 10 GeV the differential cross section behaved 
there differently than shown by Eq. (\ref{inter}) at 13 TeV, namely,
\begin{equation}
d\sigma /dt \propto \exp [-b\sqrt {\vert t \vert }] \approx \exp [-bp\theta ].
\label{int10}
\end{equation}
That was called the Orear regime by the name of its discoverer \cite{orear}.
The $\sqrt {\vert t\vert }$-exponential behavior outside the diffraction cone
replaced the $t$-exponential one inside it.

However, the experimental data at 13 TeV show that the above 
$\theta$-exponential regime at intermediate transferred momenta outside the 
diffraction cone changes drastically again to the much faster $t$-exponential 
($\theta $-Gaussian) decline of Eq. (\ref{inter}). It is similar to the 
behavior in the diffraction cone but with the much smaller exponent. If
interpreted in terms of the spatial sizes, it looks as if another smaller 
size starts playing a role. This puzzle seems especially 
intriguing because the $\sqrt {\vert t\vert}$-dependence of Eq. (\ref{int10}) 
was derived as a consequence of the unitarity condition \cite{acs, copa, andr}
and of the multiple Pomeron exchange \cite{andy}. Then the general question 
arises of what happens with the unitarity and what mechanism is at work 
at 13 TeV.

The unitarity condition $SS^+=1$ imposed on the $S$-matrix can be written in 
terms of the scattering amplitudes as \cite{andr}
\begin{eqnarray}
{\rm Im} A(p,\theta )&=&\frac {1}{32\pi ^2}\int d\cos \theta _1 d\cos \theta _2 
\frac {{\rm Re} A(p,\theta _1){\rm Re} A(p,\theta _2)+
{\rm Im} A(p,\theta _1){\rm Im} A(p,\theta _2)} 
{\sqrt {[\cos \theta -\cos (\theta _1+\theta _2)]
[\cos (\theta _1-\theta _2)-\cos \theta]}}\nonumber \\
&+&F(p,\theta )=
\int d\Phi _2AA^*+\Sigma _n\int d\Phi _nM_n(p,0)M^*_n(p,\theta ).
\label{unit}
\end{eqnarray}
The first terms in both sums denote the contributions from purely elastic
rescattering to the imaginary part of the elastic amplitude ${\rm Im} A$. The
second terms (called the overlap function $F$ \cite{hove}) correspond to the
interference of the inelastic amplitudes $M_n$ of $n$-particle production
with initial and final two-particle states turned at the angle $\theta $. 
The integration region is given by the conditions
\begin{equation}
\vert \theta _1-\theta _2\vert \leq \theta ; \;\;
\theta \leq \theta _1+\theta _2 \leq 2\pi -\theta .
\label{integ}
\end{equation}
The unitarity condition (\ref{unit}) is usually considered with two assumptions
that the contributions of the real parts to the integral term and the overlap 
function are small compared to the role of imaginary parts in the integral.
First assumption stems from smallness of ${\rm Re}A$ in the diffraction cone.
It is small compared to ${\rm Im}A$ in the forward direction \cite{drna} and 
possesses zero within the cone \cite{mart}. Intuitively, the overlap 
function is also small because
inelastic processes at high energies proceed within narrow cones along the
colliding particles. Therefore the overlap of these cones must be small since 
the two-particle states of the overlap function in the unitarity condition 
are considered at the large transferred momentum. This is indicated by the 
arguments 0 and $\theta $ of the inelastic matrix elements $M_n$ which
show the initial head-on collision leading to the final two-particle state
turned at the angle $\theta $. 

If these assumptions are accepted and the corresponding terms are omitted,
the unitarity condition becomes the non-linear integral equation for the
imaginary parts of the amplitude ${\rm Im}A$. Its iterative solutions were 
attempted \cite{acs, copa}. They lead to the Orear regime. However, the obtained
values of the exponent did not agree with experimental ones. Too many iterations 
were required so that the iteration series started to contradict the unitarity.
The model with many Pomeron exchanges was also exploited \cite{andy}. It also 
lead to the qualitatively correct Orear behavior albeit with somewhat different 
main exponent and additional (unobserved!) oscillations.

However, the non-linear equation can be actually transformed to the linear one
\cite{andr}. It is possible because the main contribution to the integral term
comes from asymmetrical angles. Thus, one of ${\rm Im}A$ can be inserted from
the diffraction cone replacing it by $\sqrt {d\sigma /dt}$ from 
Eq. (\ref{diff}) while another one is at the angle close to $\theta $.
The final solution is of the Orear type. The exponents happened to be close to
their experimental values at energies about 10 GeV \cite{andr,adg}. Even though 
the direct iterative approach is not necessary here, one is tempted to consider 
the Orear regime as a consequence of some (may be, finite number) iterations, 
i.e., as rescatterings induced by the external layers of protons.

Surely, the unitarity should be valid at any energy. The $t$-exponential
decrease of the differential cross section at 13 TeV can be explained by
the new mechanism of scattering at intermediate transferred momenta.
Protons are able to penetrate inside each other deeper at higher energies
and higher transferred momenta. Thus in place of the multiple rescattering of
the external layers in the GeV-energy range the double scattering of the 
external and deeper layers happens at the TeV energies. Both scatterings
are Gausian ones in terms of the angles with exponents corresponding to two
different internal sizes.  In the unitarity relation (\ref{unit}) one of 
${\rm Im}(p,\theta _i)$ should be used as a $\theta _i$-Gaussian exponent 
with the size of the external layer $B$ and another one as a 
$\theta _i$-Gaussian exponent with the size of the internal layer $\beta $.
In experiment, one would observe the $t$-exponential decrease of the 
differential cross section with the exponent
\begin{equation}
C=\frac {B\beta}{B+\beta}.
\label{exp}
\end{equation}
Using the experimental values of $B=20.4\;$GeV$^{-2}$ and $C=4.3\;$GeV$^{-2}$ 
at 13 TeV (see Fig. 1) one can find the exponent of the internal layer 
$\beta \approx 5.4\;$GeV$^{-2}$. In terms of the spatial size the internal
layer is concentrated at the radius near 0.45 fm, twice smaller
than the external size. Thus we claim that the
scattering at 13 TeV reveals the second layer inside the protons.

It is claimed in Ref. \cite{tot} that at the largest
transferred momenta from 2.1 GeV$^2$ to 4 GeV$^2$ the power law favored by the
quark counting rules can be adopted with the exponent of the order of 10.
However, the measured angles are less than 3$\cdot $10$^{-4}$, i.e., too small
for these rules to be applicable there. Moreover, the exponential fit 
(\ref{inter}) works well in a wider range from 0.7 GeV$^2$ to 3.8 GeV$^2$.

What concerns the unitarity condition, the validity of the assumptions about 
the overlap function and the contribution of the real parts of the elastic
amplitude can not be proved if only experimental data are available.
They provide $d\sigma /dt \propto ({\rm Re}A)^2+({\rm Im}A)^2$ but not the
real and imaginary parts separately. Some theoretical help is necessary
to get the forward values of the real and imaginary parts of the amplitude and 
the guesses about the zero value of the real part inside the diffraction cone.
Additional models and approximations are needed for more detailed
description of the amplitudes. The good fits of experimental data and knowledge 
of the energy behavior of real and imaginary parts of the elastic scattering 
amplitude separately are claimed in \cite{kfk, cps}. That can be used to verify
 the validity of the assumptions used for the above treatment of the unitarity 
 relation. The work is in progress.
 
Concluding, we argue that at higher energies the deeper layers of protons
enter the game for protons scattered outside the diffraction cone. 
The multiple scattering inside the external layer observed there at 
GeV-energies is replaced by a common effect of the double scattering due to 
the external and internal layers at TeV-energies. That restores the
$t$-exponential decrease of the differential cross section outside the
diffraction cone at TeV-energies which replaces the
 $\sqrt {\vert t\vert }$-exponential Orear-behavior at GeV-energies.

\vspace{0.5cm}                       

{\bf Acknowledgement}

I am grateful for support by the RAS-CERN program.


\begin{thebibliography}{99}
\bibitem{tot}
G. Antchev, P. Aspell, I. Atanasov et al. (TOTEM collaboration) arXiv:1812.08283
\bibitem{cso}
T. Cs\"{o}rg\"o, Talk at Low-x 2017, 12-18 June 2017. Bislegli. Available
online: https://indico.cern.ch/event/609299/
\bibitem{dr18}
I.M. Dremin, Physics-Uspekhi {\bf 61} (2018) 381
\bibitem{dr19}
I.M. Dremin, Particles {\bf 2} (2019) 57
\bibitem{drch}
I.M. Dremin, D.S. Chernavsky, JETP {\bf 38} (1960) 229
\bibitem{isl}
M.M. Islam, Nucl. Phys. B {\bf 104} (1976) 511
\bibitem{isku}
M.M. Islam, J. Kaspar, R.J. Luddy, Mod. Phys. Lett. A {\bf 24} (2009) 485
\bibitem{123}
V.D. Burkert, L. Elouadrhiri, F.X. Girod, Nature {\bf 557} (2018) 396
\bibitem{dr17}
I.M. Dremin, Physics-Uspekhi, {\bf 58} (2015) 61
\bibitem{cps}
T. Cs\"{o}rg\"o, R. Pasechnik, A. Ster, Acta Phys. Pol. B Proc. Suppl. 
{\bf 12} (2019) 779
\bibitem{orear}
J. Orear, Phys. Rev. Lett. {\bf 12} (1964) 112
\bibitem{acs} 
D. Amati, M. Cini, A. Stanghellini, Nuovo Cim. {\bf 30} (1963) 193
\bibitem{copa}
V.N. Cottingham, R.F. Peierls, Phys. Rev. {\bf 137} (1965) B147
\bibitem{andr}
I.V. Andreev, I.M. Dremin, JETP Lett. {\bf 6} (1967) 262
\bibitem{andy}
A.A. Anselm, I.T. Dyatlov, Phys. Lett. B {\bf 24} (1967) 479
\bibitem{hove}
L. Van Hove, Nuovo Cim. {\bf 28} (1963) 798
\bibitem{drna}
I.M. Dremin, M.T. Nazirov, JETP Lett. {\bf 37} (1983) 198
\bibitem{mart} 
A. Martin, Lett. Nuovo Cim. {\bf 7} (1973) 811
\bibitem{adg}
I.V. Andreev, I.M. Dremin, I.M. Gramenitsky, Nucl. Phys. B {\bf 10} (1969) 137
\bibitem{kfk}
A.K. Kohara, E. Ferreira, T. Kodama, Eur. Phys. J. C {\bf 74} (2014) 3175

\end{thebibliography}
\end{document}